\documentclass[useAMS,usenatbib]{mnras}

\usepackage{mathptmx}
\usepackage{mathtools}
\usepackage[T1]{fontenc}
\usepackage{txfonts}
\usepackage{ae,aecompl}
\usepackage{soul}
\usepackage{array}
\usepackage{amssymb}
\usepackage{pdflscape}

\usepackage[T1]{fontenc}
\usepackage{ae,aecompl}
\usepackage{soul}
\usepackage{indentfirst}
\usepackage{array}
\usepackage{times}

\usepackage{float}
\usepackage{graphicx}	% Including figure files

\usepackage{amsmath}	% Advanced maths commands
\usepackage{amssymb}	% Extra maths symbols
\usepackage{threeparttable}
\usepackage{subfigure}
\usepackage[all]{hypcap}

\usepackage[usenames]{color}
\usepackage{pdflscape}
\def\mnras{MNRAS}
\def\apjl{ApJL }

\def\apj{ApJ }

\def\pasp{PASP }

\def\apjs{ApJS }
\def\araa{ARAA }
\def\aap{A\&A }

\hypersetup{colorlinks=true,citecolor=blue,linkcolor=blue,filecolor=black,runcolor=black,breaklinks=true}
\definecolor{purple}{RGB}{160,32,240}
\definecolor{Cerulean}{RGB}{0,123,167}
\usepackage{etoolbox}

\newcommand{\mbh}{$M_{\bullet}$}

\newcommand{\mstar}{$M_{*}$}

\newcommand{\Msun}{M_{\odot}}

\newcommand{\bhbm}{$M_\bullet$--$M_\mathrm{bulge}$}
\newcommand{\bmsm}{$M_\mathrm{bulge}$--$M_*$}
\newcommand{\bhsm}{$M_\bullet$--$M_*$}

\newcommand{\jwst}{\textit{JWST}}

\newcommand{\atacamaarray}{\textit{Atacama Large Millimeter Array}}
\newcommand{\alma}{\textit{ALMA}}

\title[\textsc{Trinity}: The Luminosity Bias of \bhsm{} relation at $z=6$]{\textsc{Trinity} II: The Luminosity-dependent Bias of the Supermassive Black Hole Mass--Galaxy Mass Relation for Bright Quasars at z=6}

\author[H. Zhang et al.]{
Haowen Zhang,$^{1}$\thanks{E-mail: hwzhang0595@email.arizona.edu}
Peter Behroozi,$^{1}$
Marta Volonteri,$^{2}$
Joseph Silk,$^{2,3,4}$
\newauthor{
Xiaohui Fan,$^{1}$
Philip F. Hopkins,$^{5}$
Jinyi Yang,$^{1,\ast}$
and James Aird$^{6,7}$
}
\\
$^{1}$University of Arizona, 933 N Cherry Ave., Tucson, AZ 85721, USA, \\
$^{2}$Institut d'Astrophysique de Paris (UMR 7095: CNRS \& Sorbonne Universite), 98 bis Bd. Arago, F-75014, Paris, France\\
$^{3}$Department of Physics and Astronomy, Johns Hopkins University, Baltimore, MD 21218, USA\\
$^{4}$BIPAC, Department of Physics, University of Oxford, Keble Road, Oxford OX1 3RH, UK\\
$^{5}$Theoretical Astrophysics, California Institute of Technology, Pasadena, CA 91125, USA\\
$^{6}$Institute for Astronomy, University of Edinburgh, Royal Observatory, Edinburgh EH9 3HJ, UK\\
$^{7}$Department of Physics and Astronomy, University of Leicester, University Road, Leicester LE1 7RH, UK\\
$^{\ast}$Strittmatter Fellow
}

\date{Accepted XXX. Received YYY; in original form ZZZ}

\pubyear{2021}

\begin{document}
\label{firstpage}
\pagerange{\pageref{firstpage}--\pageref{lastpage}}
\maketitle

\begin{abstract}
Using recent empirical constraints on the dark matter halo—galaxy—supermassive black hole (SMBH) connection from $z=0-7$, we infer how undermassive, typical, and overmassive SMBHs contribute to the quasar luminosity function (QLF) at $z=6$. We find that beyond $L_\mathrm{bol} = 5\times 10^{46}$ erg/s, the $z=6$ QLF is dominated by SMBHs that are at least 0.3 dex above the $z=6$ median \bhsm{} relation. The QLF is dominated by typical SMBHs (i.e., within $\pm 0.3$ dex around the \bhsm{} relation) at $L_\mathrm{bol} \lesssim 10^{45}$ erg/s. At $z\sim 6$, the intrinsic \bhsm{} relation for all SMBHs is slightly steeper than the $z=0$ scaling, with a similar normalization at $M_* \sim 10^{11} M_\odot$. We also predict the \bhsm{} relation for $z=6$ bright quasars selected by different bolometric luminosity thresholds, finding very good agreement with observations. For quasars with $L_\mathrm{bol} > 3\times 10^{46}$ ($10^{48}$) erg/s, the scaling relation is shifted upwards by $\sim 0.35$ (1.0) dex for $10^{11}\Msun$ galaxies. To accurately measure the intrinsic \bhsm{} relation, it is essential to include fainter quasars with $L_\mathrm{bol} \lesssim 10^{45}$ erg/s. At high redshifts, low-luminosity quasars are thus the best targets for understanding typical formation paths for SMBHs in galaxies.
\end{abstract}

\begin{keywords}
galaxies: haloes -- galaxies: evolution -- quasars: supermassive black holes
\end{keywords}

%%%%%%%%%%%%%%%%%%%%%%%%%%%%%%%%%%%%%%%%%%%%%%%%%%

%%%%%%%%%%%%%%%%% BODY OF PAPER %%%%%%%%%%%%%%%%%%

\vspace{-1.5cm}

\section{Introduction}
\label{s:introduction}

The supermassive black hole (SMBH) masses of high-redshift quasars contain critical information on 1) the formation and growth of SMBHs at high redshifts; 2) the feedback from active SMBHs (also called active galactic nuclei, AGN) on their host galaxies in the early Universe, and 3) the build-up of the galaxy--SMBH mass connection. Consequently, there have been many high-redshift quasar surveys aimed at studying their demography. Currently, there are $275$ quasars known at $z > 6$ \citep{Fan2023}. Due to the sheer brightness of high-redshift quasars, it is impractical to measure their host galaxy properties by fitting galaxy spectral energy distributions (SEDs). Therefore, galaxy dynamical masses are often used as a proxy for stellar masses. The measurement of galaxy dynamical mass relies on the high spatial resolution and sensitivity of interferometric radio observations. As a result, the existing galaxy mass measurements have been made predominantly by the \atacamaarray{} (\alma{}; see the compilation by \citealt{Izumi2021} and references therein). With the launch of \jwst{}, we are finally able to measure host galaxies' stellar masses from rest-frame optical light (e.g., \citealt{Ding2022}) for at least some high-redshift quasars. At face value, these quasars seem to lie well above the local SMBH mass--galaxy mass (\bhsm{}) relation, i.e., having overmassive SMBHs relative to those typical for $z=0$ galaxies. However, this higher observed \bhsm{} relation at $z=6$ can result from systematic effects. Specifically, the bright quasar sample may be biased towards overmassive SMBHs when there is scatter around the intrinsic \bhsm{} scaling relation. These quasars are often selected using flux-limited photometric surveys in the optical and infrared wavebands. When SMBHs have similar Eddington ratios, overmassive objects (compared to the median \bhsm{} relation) would be brighter, and will be overrepresented in the selected sample (also known as Lauer bias; \citealt{Lauer2007}). With a given intrinsic \bhsm{} relation and Eddington ratio distribution, the magnitude of Lauer bias increases with the scatter in \mbh{} at fixed \mstar{}, since larger scatter leads to more overmassive SMBHs in the quasar sample. In the absense of  scatter around the \bhsm{} relation, there will be \emph{no} such selection bias, because every single quasar in the sample will lie perfectly on the scaling relation. To estimate the extent of Lauer bias, one thus needs: 1) the scatter around the intrinsic \bhsm{} relation; and 2) the underlying Eddington ratio distributions for SMBHs in different galaxies (see, e.g., \citealt{Li2022}).

In this work, we measure the effect of selection bias on the \bhsm{} relation for $z\sim6$ quasars with \textsc{Trinity} \citep[][]{Zhang2021}. \textsc{Trinity} is an empirical model of the dark matter halo—galaxy—SMBH connection from $z=0-10$. With joint constraints from galaxy observations from $z=0-10$ and SMBH observations from $z=0-6.5$, \textsc{Trinity} reconstructs consistent SMBH growth histories and Eddington ratio distributions, both of which are functions of halo/galaxy mass and redshift. This information enables us to create mock luminosity-selected quasar samples and directly compare their \bhsm{} relations with the intrinsic relation for \emph{all} $z\sim 6$ SMBHs. This work is timely at the beginning of the \jwst{} era, because our results will: 1) predict the offset in the observed \bhsm{} relation vs.\ the intrinsic relation, as a function of quasar luminosity; 2) quantify the extent to which pure selection bias can explain the apparent redshift evolution in the \bhsm{} relation from $z=0$ to $z=6$; and 3) point future \jwst{} observations towards better quasar samples for more accurate measurement of the \bhsm{} relation at high redshifts. These predictions are directly testable by future \jwst{} observations.

The paper is organized as follows. \S \ref{s:method} covers methodology. In \S \ref{s:sims_and_data}, we describe the dark matter simulation and galaxy/SMBH observations used to constrain \textsc{Trinity}.  \S \ref{s:results} presents our findings on the quasar mass/luminosity bias at $z=6$. Finally, we present conclusions in \S \ref{s:conclusions}.  In this work, we adopt a flat $\Lambda$CDM cosmology with parameters ($\Omega_m=0.307$, $\Omega_{\mathrm{\Lambda}}=0.693$, $h=0.678$, $\sigma_8=0.823$, $n_s=0.96$) consistent with \textit{Planck} results \citep{Planck2016}. We use datasets that adopt the Chabrier stellar initial mass function \citep[IMF, ][]{Chabrier2003}, the \citet{Bruzual2003} stellar population synthesis model, and the Calzetti dust attenuation law \citep{Calzetti2000}. Halo masses are calculated following the virial overdensity definition from \citet{Bryan1998}. 

\vspace{-0.5cm}

\section{Methodology}
\label{s:method}

% In \S\ref{ss:why_trust_trinity} we explain: 1) why \textsc{Trinity} is able to infer the halo--galaxy--SMBH connection from $z=0-10$ uniquely, robustly, and self-consistently; and 2) why such inference can be done by fitting observations alone, without the need to assume any particular physical mechanisms for, e.g., halo--galaxy--SMBH interactions. \S\ref{ss:overview} gives an overview of the \textsc{Trinity} implementation.

% \vspace{-0.5cm}

\subsection{Why observations alone can constrain the halo--galaxy--SMBH connection}
\label{ss:why_trust_trinity}

The So\l{}tan argument \citep[][]{Soltan1982} gave rise to the earliest empirical models of SMBH growth: the ratio of the total luminosity output of SMBHs to their $z=0$ mass density gives the cosmic average radiative efficiency (see, e.g., \citealt{Yu2002,Marconi2004}). This in turn allows inferring the cosmic average growth history of SMBHs from the redshift evolution of the total luminosity in QLFs.

Recently, studies including \citet[][]{Yang2018} and \citet[][]{Aird2018} have measured quasar luminosity distributions \emph{as functions of host galaxy mass}. At the same time, empirical models of the halo--galaxy connection have succeeded in reconstructing robust galaxy assembly histories that are constrained by galaxy data from $z=0-10$ (e.g., \citealt{Behroozi2013,Moster2013,Moster2018,Behroozi2019}). These breakthroughs enabled, e.g., \citet[][]{Shankar2020} and \citet[][]{Zhang2021}, to apply the So\l{}tan argument to galaxies split into different stellar mass bins. Specifically, the cumulative SMBH mass growth of a chosen galaxy population is proportional to the net SMBH luminosity of the galaxies' progenitors.  This luminosity (of the SMBH progenitors) may be measured by combining measured SMBH luminosities for the correct distribution of galaxy progenitor masses (as a function of redshift), where the galaxy progenitor mass distribution is given by the above-mentioned constraints on galaxy growth histories. The radiative efficiency (which allows inferring the SMBH growth history) is then given by the ratio of the galaxies' net SMBH progenitor luminosity to the galaxies' $z=0$ SMBH masses. Applying the So\l{}tan argument in this way yields simultaneous growth histories of galaxies and SMBHs, and in particular constrains the evolution of the SMBH mass -- galaxy mass relation with redshift. Based on \textsc{Trinity}'s predicted SMBH growth histories in different halo/galaxy populations, we modeled their mass and Eddington ratio distributions, which are constrained by SMBH observations, e.g., quasar luminosity distributions as functions of stellar mass, and total quasar luminosity functions.

\textsc{Trinity} also explicitly models the scatter around the median \bhsm{} relation, which is constrained by the shape of active SMBH mass functions. With inferred SMBH Eddington ratio distributions and the \bhsm{} scatter, \textsc{Trinity} is well positioned to predict Lauer bias. Constraints on the Lauer bias come from comparing active SMBH mass functions to the expected total SMBH mass function arising from the SMBH--galaxy relationship constrained above, as well as measured Eddington ratios for bright quasars. Stronger Lauer bias results in overmassive black holes being more likely to be active; similarly, stronger Lauer bias also results in lower Eddington ratios at fixed luminosity (as only the most massive black holes are then allowed to be the most luminous). Quantitatively, we find that SMBHs with different \mbh{} (at fixed galaxy mass) have nearly identical Eddington ratio distributions at z $\sim$ 6. This is constrained by the fact that there are very few observed low-mass quasars with super-Eddington accretion at these redshifts (e.g., \citealt{Shen2019}).

Finally, we have verified that \textsc{Trinity} predictions are robust against changes in model parameterizations and input assumptions. We experimented with many model variants by changing, for example: 1) the way to parametrize the \bhsm{} relation; 2) Eddington ratio distribution shapes; 3) SMBH merger prescriptions; 4) AGN obscuration corrections; 5) AGN bolometric corrections; we found no qualitative change in our predictions, when the input observations were self-consistent with each other. For full details, we refer readers to the Appendices of \citet[][]{Zhang2021}.

\vspace{-0.5cm}

\subsection{Implementation overview}
\label{ss:overview}

Here, we give a brief overview of \textsc{Trinity}. For full details, we refer readers to \citet{Zhang2021}.

\textsc{Trinity} parameterizes the halo—galaxy connection similarly to the \textsc{UniverseMachine}: the galaxy star formation rate (SFR) is a double-power law of the galaxy's peak halo mass, and the fraction of star-forming galaxies is a sigmoid function of halo mass. Both functions are allowed to evolve with redshift, which are constrained by galaxy datasets. This parameterization has been well-tested in \citet{Behroozi2019}, and gives robust inference of the halo--galaxy connection from joint observational galaxy constraints.

We make the galaxy--SMBH connection in \textsc{Trinity} by parameterizing the \bhbm{} relation as a redshift-dependent power-law. To convert galaxy \emph{total} masses into bulge masses, we use a redshift-dependent \bmsm{} scaling relation that is fit to SDSS and CANDELS observations (for full details, see \citealt{Zhang2021}). We calculate average SMBH growth rates in different halo/galaxy populations by tracking the change in typical \mbh{} between successive snapshots. We then convert average SMBH growth rates into AGN Eddington ratio distributions with the following AGN properties chosen from the parameter space: 1) fractional contributions from SMBH accretion vs. SMBH mergers; 2) the correlation between SMBH mass and SMBH accretion rate at fixed halo mass, $\rho_\mathrm{BH}$; 3) the AGN energy efficiency; 4) the AGN duty cycle; and 5) Eddington ratio distribution shapes. With SMBH masses and Eddington ratio distributions fully parameterized, we generate SMBH observables including quasar luminosity functions, quasar luminosity distributions as functions of host galaxy mass, active SMBH mass functions, and the $z=0$ \bhbm{} relation. We include systematic and selection effects such as AGN obscuration and bolometric corrections, and finally compare the generated statistics with observed data.

Using a custom Metropolis Markov Chain Monte Carlo (MCMC) algorithm (based on \citealt{Haario2001}), we create $\sim 2$ million mock universes and compare them with our data compilation. Through such comparisons, we obtain the joint posterior distribution of \textsc{Trinity} model parameters, and characterize the best-fitting halo--galaxy--SMBH connection, as well as the corresponding uncertainties.

\vspace{-1.0cm}

\section{Simulations and Data Constraints}
\label{s:sims_and_data}

\subsection{Dark Matter Halo Statistics}
\label{ss:dm_sims}

\textsc{Trinity} traces statistical halo assembly histories obtained from N-body simulations of dark matter haloes, instead of keeping track of individual haloes/galaxies across cosmic time. Specifically,  halo mass functions are obtained from \citet{Tinker2008}, with the corrections in \cite{Behroozi2013} to: 1) use halo peak mass instead of current mass; 2) improve the accuracy at higher redshifts; and 3) include satellite haloes. We refer readers to Appendix G of \citet{Behroozi2013} for details. These mass functions are valid for studying halo evolution from at least $10^{10} M_\odot$ to $10^{15} M_\odot$.

Haloes experience mass growth via both accretion and mergers. The average halo accretion histories in this work are described by the fitting formulae in Appendix H of \citet{Behroozi2013}. Halo merger rates are fitted from the mock catalogs of the \textsc{UniverseMachine} (\citealt{Behroozi2019}). The fitting formulae for halo mergers are presented in Appendix A of \citet{Zhang2021}.

\vspace{-0.8cm}

\subsection{Observational Data Constraints}
\label{ss:obs_data}

We use the following galaxy data to constrain the halo--galaxy connection: stellar mass functions (SMFs, $z=0-8$), galaxy quenched fractions (QFs, $z=0-4$), average specific star formation rates (SSFRs, $z=0-8$), cosmic star formation rates (CSFRs, $z=0-10$), and galaxy UV luminosity functions (UVLFs, $z=8-10$). We refer readers to \S 2.2 and Appendix C of \citet[][]{Behroozi2019} for full details about all adopted galaxy data. 

To constrain the galaxy--SMBH connection, we have compiled the following SMBH observables: X-ray quasar luminosity functions (QLFs, from \citealt{Ueda2014}, $z=0-5$), X-ray quasar probability distribution functions (QPDFs, from \citealt{Aird2018}, $z=0.1-2.5$), optically-selected active black hole mass functions (ABHMFs, from \citealt{Schulze2010}, \citealt{Schulze2015}, and \citealt{Kelly2013}, $z=0.2-5$), the $z=0$ \bhbm{} relation \citep{Haring2004,Beifiori2012,Kormendy2013,McConnell2013,Savorgnan2016}, and the observed \mbh{} distribution of high redshift ($z\sim 6$) bright quasars \citep{Shen2019}. These SMBH data cover $z=0-6.5$. For more details about these SMBH observables, see \S 3.2.2 of \citet{Zhang2021}.

\vspace{-0.5cm}

\section{Results}
\label{s:results}

\subsection{Offsets in the \bhsm{} relation for bright quasars vs.\ all SMBHs at $z=6$}
\label{ss:bhsm_quasar_bias}

\begin{figure}
\subfigure{
\includegraphics[width=0.48\textwidth]{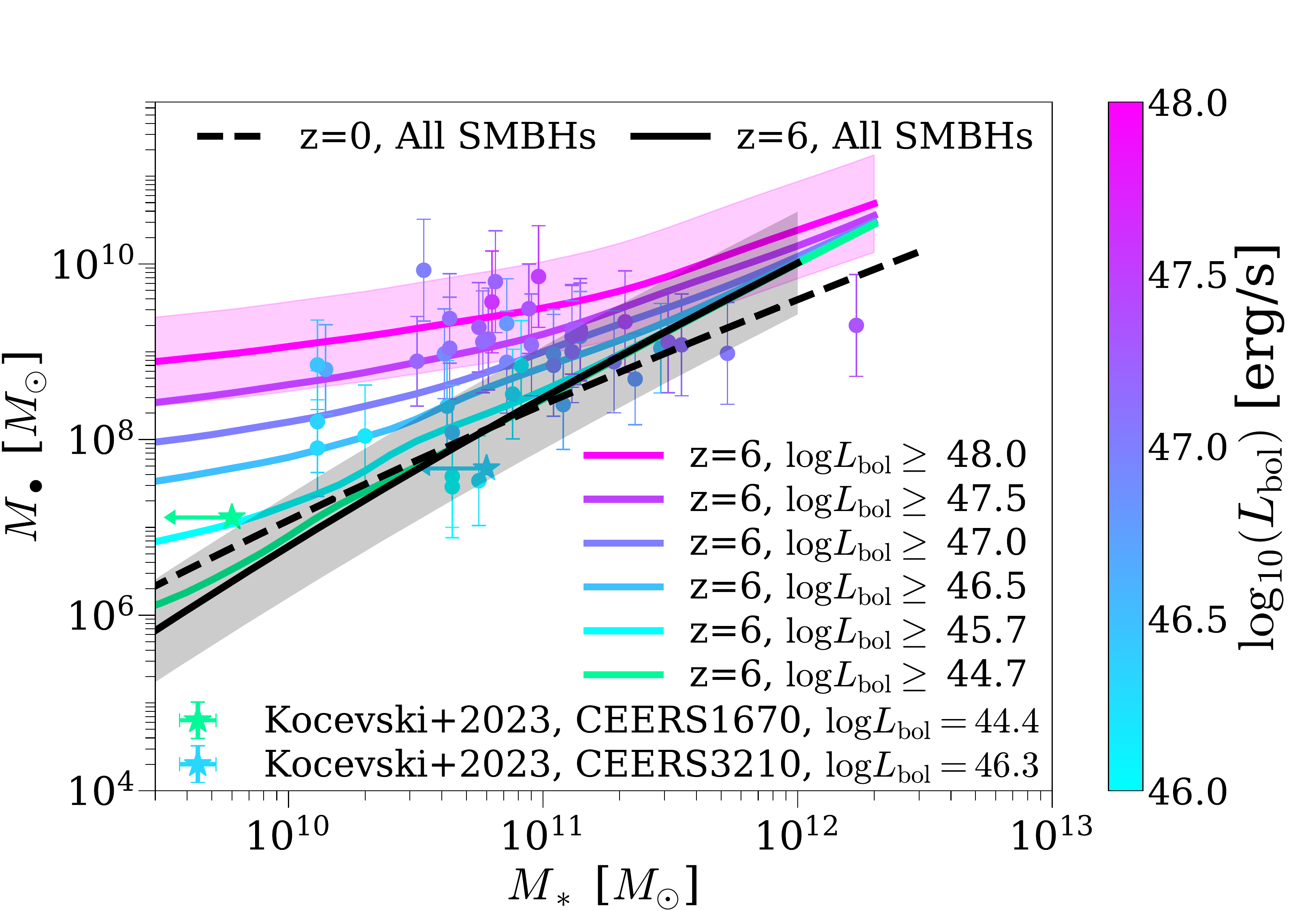}
}
% \caption{The median \bhsm{} relation for quasars with different lower limits in bolometric luminosity at $z=6$. The red shaded region is the $1-\sigma$ spread around the median scaling relation for quasars ($\sim0.55$ dex), which includes the random scatter in observed \mbh{} when using virial estimates. This \mbox{(log-)}normal scatter is nearly luminosity-independent, so we only show it for the brightest quasars for clarity. The black solid line is the \bhsm{} relation for all SMBHs at $z=6$, and the black shaded region is the intrinsic$+$observed scatter around the intrinsic \bhsm{} relation. The green solid line is the \bhsm{} relation for AGNs brighter than $\log L_\mathrm{bol} [\mathrm{erg/s}] \geq 44.7$, which is the approximate lower limit for \jwst{} to measure \mbh{} via broad emission lines. For comparison, we also show the $z=0$ relation in the black dashed line. Individual data points are $z\gtrsim 6$ quasars compiled by \citet[][]{Izumi2021}, assuming that the dynamical mass is a proxy for the total stellar mass. The two $z>5$ low-luminosity AGNs from \citet{Kocevski2023} are shown in stars. We adopted an uncertainty in \mbh{} of 0.5 dex for both AGNs. For CEERS 3210, we adopted \mbh{} and \mstar{} values based on an extinction of $A_V=4$, as is done in Fig.\ 7 of \citet{Kocevski2023}.  See \S\ref{ss:bhsm_quasar_bias} for discussion.}
\caption{The $z=6$ median \bhsm{} relation for quasars with different bolometric luminosity thresholds. The red shaded region is the $1-\sigma$ spread around the median scaling relation for quasars ($\sim0.55$ dex), which includes the random scatter in observed \mbh{} when using virial estimates. This \mbox{(log-)}normal scatter is nearly luminosity-independent, so we only show it for the brightest quasars for clarity. The black solid line is the \bhsm{} relation for all SMBHs at $z=6$, and the black shaded region is the intrinsic$+$observed scatter around the intrinsic \bhsm{} relation. The green solid line is the \bhsm{} relation for AGNs brighter than $\log L_\mathrm{bol} [\mathrm{erg/s}] \geq 44.7$, which is the approximate lower limit for \jwst{} to measure \mbh{} via broad emission lines. For comparison, we also show the $z=0$ relation in the black dashed line. Individual data points are $z\gtrsim 6$ quasars compiled by \citet[][]{Izumi2021}. The two $z>5$ AGNs from \citet{Kocevski2023} are shown in stars. See \S\ref{ss:bhsm_quasar_bias} for discussion.}

\label{f:bhsm_bias_z6_obs}
\end{figure}

\begin{figure}

\subfigure{
\includegraphics[width=0.48\textwidth]{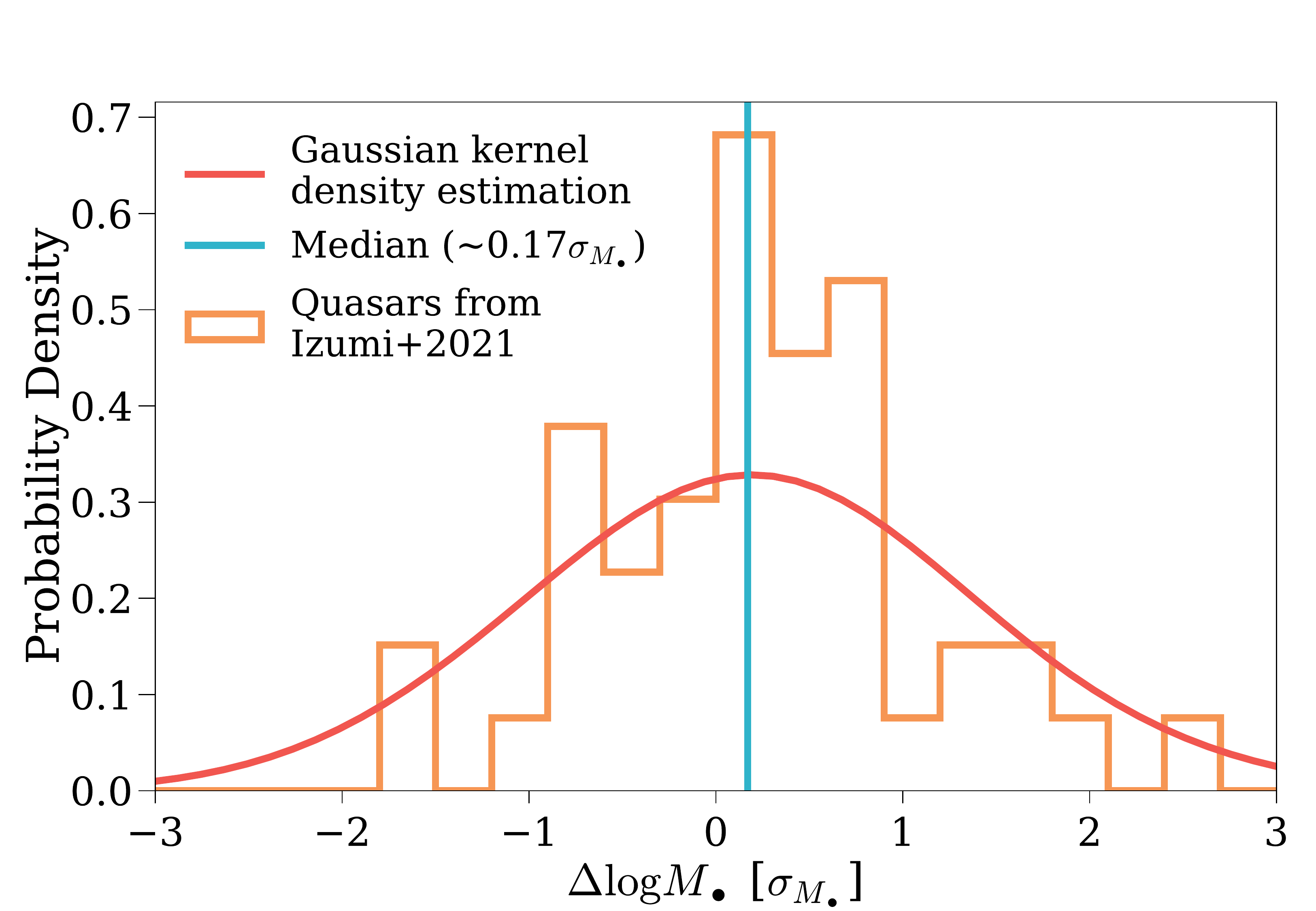}
}
\caption{The distribution of \citet[][]{Izumi2021} $z=6$ quasars' deviation from the \textsc{Trinity} \bhsm{} relations. For each quasar, we calculate the corresponding quasar \bhsm{} relation at its bolometric luminosity. The deviation is further normalized by the Gaussian spread, $\sigma_{M_\bullet}$, which is a quadratic sum of the intrinsic scatter around the \bhsm{} relation, and the uncertainty in the observed \mbh{}. The Gaussian kernel density estimation of the distribution is shown in the red curve. See \S\ref{ss:bhsm_quasar_bias}.}
\label{f:dmbh_sigma_z6}
\end{figure}

Fig.\ \ref{f:bhsm_bias_z6_obs} shows the the median \bhsm{} relation for $z=6$ quasars from \textsc{Trinity}, as a function of the lower limit in bolometric luminosity. For reference, the bolometric quasar luminosity limit is $\log L_\mathrm{bol}\gtrsim 45.5$ for \textit{Subaru} SHELLQs--Wide program \citep{Matsuoka2016,Runnoe2012a,Runnoe2012b}, and $\log L_\mathrm{bol}\gtrsim 46.5$ for Pan-STARRS1 (PS1, \citealt{Chambers2016}) and SDSS \citep{Jiang2016}. According to \textsc{Trinity}, the Eddington ratio distribution is nearly mass-independent at $z=6$, so more massive black holes are naturally brighter and more likely to be included in the sample. As a result, SMBHs in bright quasars tend to be overmassive compared to their host galaxies. This systematic offset increases with quasar luminosity, from a $\sim 0.35$ dex offset for $\log L_\mathrm{bol} [\mathrm{erg/s}]\geq 46.5$ to a $\sim 1$ dex offset for $\log L_\mathrm{bol} [\mathrm{erg/s}]\geq 48$, at host stellar masses of $M_* \sim 10^{11} M_\odot$. The shaded region denotes the 1$-\sigma$ (log-)normal spread around the median scaling relation for luminosity limited samples, which is $\sim 0.55$ dex across the mass and luminosity ranges at $z=6$. This spread includes the scatter in intrinsic \mbh{} at fixed \mstar{}, as well as the scatter in \emph{observed} \mbh{} around the intrinsic values. This spread is similar to the one around the intrinsic \bhsm{} relation for all SMBHs, which is shown in the black shaded region. This is because both spreads are dominated by the typical scatter in \emph{observed} \mbh{} around the intrinsic values ($\sim$ 0.5 dex). On the other hand, the \emph{intrinsic} scatter is slightly smaller for the biased sample than for all SMBHs, due to the selection in AGN luminosity. Qualitatively, this trend of increasing \bhsm{} normalization with higher luminosity is consistent with the observations, such as the data points compiled by \citet{Izumi2021} (colour-coded by bolometric luminosity). We converted the rest-frame $1450\AA$ magnitudes from \citet{Izumi2021} into bolometric luminosities using the bolometric correction from \citet{Runnoe2012a,Runnoe2012b}. In Fig.\ \ref{f:bhsm_bias_z6_obs}, we also show the two $z>5$ low-luminosity AGNs from \citet{Kocevski2023} with star-shaped points. The galaxy masses of both AGNs are estimated with SED fitting ignoring potential contributions from their AGNs, and thus should be treated as upper limits. Taken at face value, the \mbh{}/\mstar{} ratios of these two AGNs are qualitatively consistent with \textsc{Trinity}'s predictions. However, further follow-up observations are required for a better measurement of host galaxy masses.
 
Fig.\ \ref{f:dmbh_sigma_z6} shows the deviation in \mbh{} from the \textsc{Trinity} \bhsm{} relations for the \citet[][]{Izumi2021} quasar sample. For each observed quasar with a bolometric luminosity $L_0$, we calculate the \textsc{Trinity} \bhsm{} relation for quasars with $\log L_0 - 0.1\ \mathrm{dex} < L_\mathrm{bol} < \log L_0 + 0.1\ \mathrm{dex}$ to ensure a fair comparison. The \mbh{} deviation is divided by the (log-normal) standard deviation $\sigma_{M_\bullet}$, which is the quadratic sum of the intrinsic \textsc{Trinity} \bhsm{} scatter and the measurement uncertainty in \mbh{} from \citet[][]{Izumi2021}. The distribution of the \mbh{} deviations has a significant amount of scatter around the median value of $\sim 0.17\sigma_{M_\bullet}$, which is $\lesssim 0.1$ dex. Therefore, the apparent evolution in the \bhsm{} relation from $z=0$ to $z=6$ can be largely explained by Lauer bias. In the future, more accurate and precise measurements of SMBH and galaxy masses (stellar masses from, e.g., \jwst{} and dynamical masses from \alma{}) are needed to understand the slight positive deviation as shown in Fig.\ \ref{f:dmbh_sigma_z6}.

According to \textsc{Trinity}, there is only mild evolution in \bhsm{} from $z=0$ to $z=6$. This means that \emph{typical} SMBHs on the intrinsic \bhsm{} relation \emph{do not} experience significant mass build-up before their host galaxies, even though it may be the case for \emph{overmassive} (and thus brighter) SMBHs in current quasar samples. To understand typical SMBHs and host galaxies' growth histories, it is thus essential to measure the \bhsm{} relation of less biased (i.e., fainter) quasar samples. In Fig.\ \ref{f:bhsm_bias_z6_obs}, we also show the median \bhsm{} relation for all SMBHs brighter than $\log L_\mathrm{bol} [\mathrm{erg/s}] > 44.7$, which is the lowest AGN luminosity at which \jwst{} can still measure \mbh{} reasonably well. At such a low luminosity threshold, the observed \bhsm{} relation is very close to the intrinsic relation for all the SMBHs at $\log M_* \gtrsim 10.5$. Therefore, to accurately measure the $z\sim 6$ \bhsm{} relation without a severe selection bias, it is essential to focus on fainter quasars at $\log L_\mathrm{bol} [\mathrm{erg/s}] \lesssim 45$ in $\log M_* \gtrsim 10.5$ galaxies (e.g., those detected in the Subaru High-z Exploration of Low-Luminosity Quasars (SHELLQs), \citealt{Matsuoka2022}). This is also in line with a series of theoretical studies with Monte Carlo and hydro-dynamical simulations, e.g.,\citet{Volonteri2011,Volonteri2016b,Marshall2020,Habouzit2022}. We do caution that for fainter quasars with $\log L_\mathrm{bol} [\mathrm{erg/s}] \lesssim 45$, the increasing scatter in the bolometric correction at a fixed UV luminosity for individual objects (see, e.g., \citealt{Runnoe2012a,Runnoe2012b}) could add additional uncertainties to bolometric luminosity estimates. This may complicate the interpretation of the \bhsm{} relation for fainter quasars in the future.

In addition to the random scatters around intrinsic \mbh{} values, the observed \mbh{} are also subject to potential systematic offsets. Such offsets could caused by different reasons, e.g., the use of the same virial estimate calibration at both low and high redshifts, which may not be accurate in the real Universe. Qualitatively, if the observed \mbh{} values are systematically overestimated(underestimated), correcting such offsets will lead to better(worse) agreement between \textsc{Trinity}'s predictions and the observations.

\vspace{-0.8cm}

% Since part of the accreted mass is still converted into radiation in the low-Eddington ratio regime, part of such BHAR  In both models, SMBHs 

\subsection{Quasar luminosity functions binned by offset from the \bhsm{} relation}
\label{ss:qlf_mstar_bias}
\begin{figure}
\subfigure{
\includegraphics[width=0.48\textwidth]{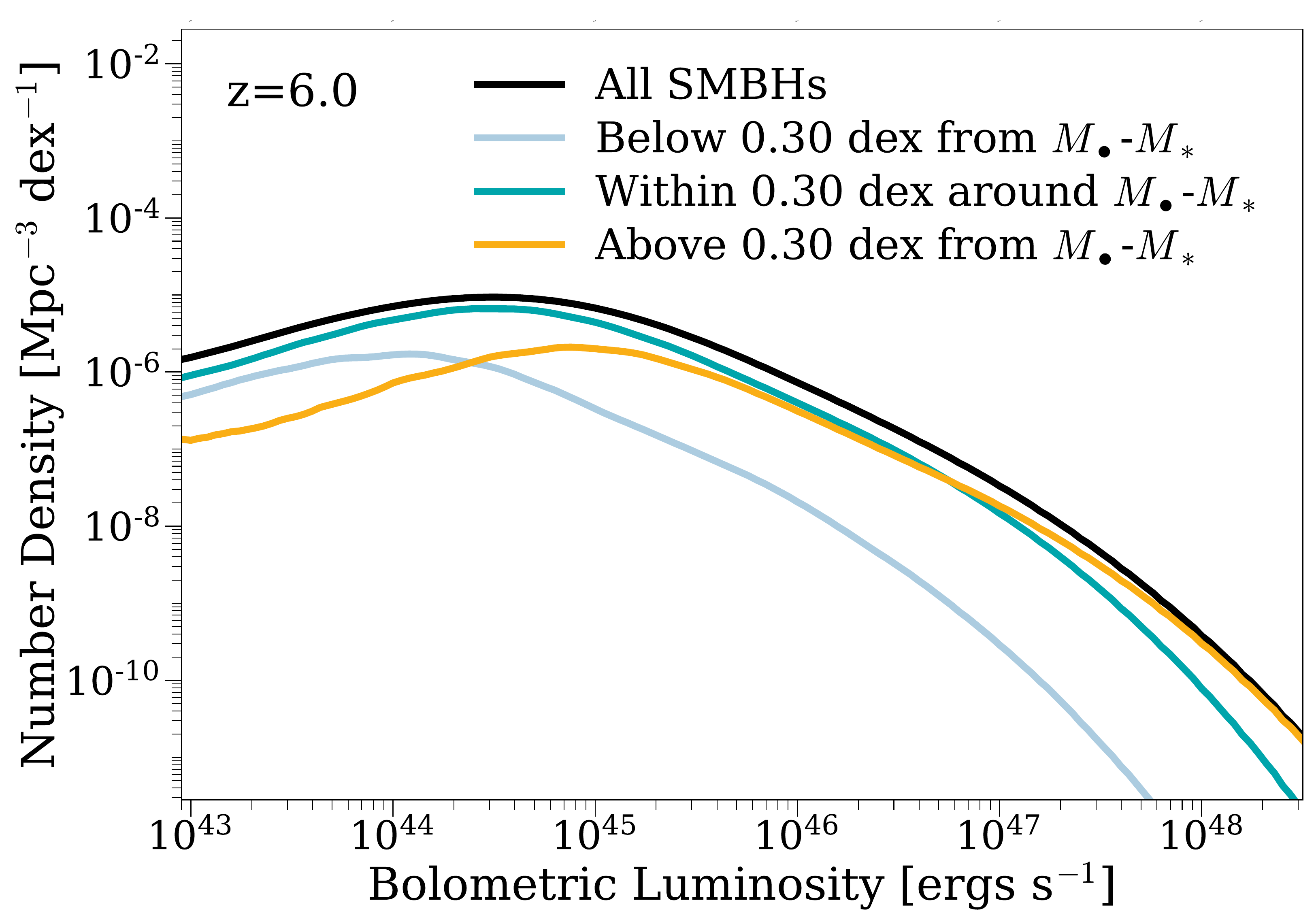}
}
\caption{Quasar luminosity functions in bins of the deviation in SMBH mass from the median \bhsm{} relation at $z=6$. See \S\ref{ss:qlf_mstar_bias}}
\label{f:qlf_mstar_bias}
\end{figure}

Fig.\ \ref{f:qlf_mstar_bias} shows QLFs in bins of offset in SMBH mass compared to the median \bhsm{} relation at different redshifts. Compared to Fig.\ \ref{f:bhsm_bias_z6_obs}, Fig.\ \ref{f:qlf_mstar_bias} quantifies the Lauer bias in another way, i.e., the amount by which brighter quasars are more likely to be driven by over-massive SMBHs (compared to their host galaxy mass) than typical SMBHs. This effect arises mainly because extremely massive host galaxies are very rare by themselves, and cannot account for the number density of high-mass black holes. According to \textsc{Trinity}, over-massive SMBHs ($>0.3$ dex above the median \bhbm{} relation) dominate QLFs at $\log_{10}L_\mathrm{bol} [\mathrm{erg/s}]\gtrsim 46.7$ and $z\sim 6$. Typical SMBHs within 0.3 dex around the intrinsic \bhsm{} relation dominate the quasar luminosity function at $\log L_\mathrm{bol} [\mathrm{erg/s}]\lesssim 45$. This quantitatively demonstrates the necessity of including fainter AGNs when comparing the \bhsm{} relations from the local Universe vs.\ $z\sim 6$ SMBHs.

\vspace{-0.7cm}

\section{Conclusions}
\label{s:conclusions}

In this work, we examine the systematic bias in the observed \bhsm{} relation for luminosity-limited quasar samples, as well as contributions to the $z= 6$ quasar bolometric luminosity function from SMBHs at different offsets relative to the $z= 6$ \bhsm{} relation. Compared to previous studies like \citet[][]{Li2022} that adopt empirically determined Eddington ratio distributions and intrinsic \bhsm{} relations at $z=0$, we make inferences based on the joint SMBH mass--Eddington ratio distributions at different redshifts from \textsc{Trinity}, which are: 1) explicitly constrained by galaxy and SMBH data (\S\ref{s:sims_and_data}), and 2) self-consistent with the reconstructed SMBH growth histories. Our key findings are:

\begin{itemize}

    \item At $z\sim 6$, the \bhsm{} relation for bright quasars selected by bolometric luminosity ($L_\mathrm{bol}$) is significantly higher than the intrinsic relation for \emph{all} SMBHs. This is because there is scatter around the intrinsic \bhsm{} relation, and we can only probe the most luminous AGN, which are overmassive compared to the intrinsic \bhsm{} relation. With a luminosity threshold of $\log L_\mathrm{bol} [\mathrm{erg/s}] > 46.5$ (48), the median \mbh{} is higher by 0.35 (1.0) dex for bright quasars than for typical black holes in $M_* = 10^{11} M_\odot$ host galaxies. Fainter quasars with $\log L_\mathrm{bol} [\mathrm{erg/s}] \lesssim 45$ in $\log M_* \gtrsim 10.5$ galaxies have average \mbh{}  very close to the typical \bhsm{} relation for all (active and non-active) SMBHs. Although the detected overmassive and bright SMBHs may have grown in mass significantly before their host galaxies, this is not the case for typical SMBHs on the intrinsic \bhsm{} relation at $z=6$, for which we are not yet able to measure \mbh{}.(\S\ref{ss:bhsm_quasar_bias}, Figs.\ \ref{f:bhsm_bias_z6_obs} and \ref{f:dmbh_sigma_z6});

    \item At $z\sim 6$, our predicted luminosity-dependent \bhsm{} relation are consistent with observations compiled by \citet{Izumi2021}, which are \emph{not} in the observational constraints for \textsc{Trinity}. This further demonstrates the validity of the \textsc{Trinity} model and its predictions. (\S\ref{ss:bhsm_quasar_bias}, Figs.\ \ref{f:bhsm_bias_z6_obs} and \ref{f:dmbh_sigma_z6});
    
    \item At $z\sim 6$, most observed quasars with $L_\mathrm{bol} \gtrsim 5\times 10^{46}$ erg/s have SMBH masses $\gtrsim 0.3$ dex higher than the median \bhsm{} relation. At brighter luminosities, the quasar luminosity function is increasingly dominated by SMBHs that are over-massive compared to the median \bhsm{} relation. This is because overmassive SMBHs are brighter at similar Eddington ratios. At $\log L_\mathrm{bol} [\mathrm{erg/s}] \lesssim 45$, the QLF is dominated by typical SMBHs--i.e., those within 0.3 dex of the \bhsm{} relation. (\S\ref{ss:qlf_mstar_bias}, Fig.\ \ref{f:qlf_mstar_bias})

\end{itemize}

In summary, future observational efforts to measure the intrinsic $z\sim 6$ \bhsm{} relation should focus on fainter quasars with $\log L_\mathrm{bol} [\mathrm{erg/s}] \lesssim 45$. This motivates future observations with \jwst{}, one of the few telescopes that can measure both \mbh{} and $M_\ast$ for these faint objects. At the same time, observations of faint quasars will directly test theoretical models (including \textsc{Trinity}) and their  predictions for the high-redshift galaxy--SMBH mass connection.

\vspace{-0.5cm}

\section*{Data availability}
\label{s:data_availability}

The parallel implementation of \textsc{Trinity}, the compiled datasets (\S\ref{ss:obs_data}), and the posterior distribution of model parameters are available \href{https://github.com/HaowenZhang/TRINITY}{\textbf{online}}.

\vspace{-0.5cm}

\section*{Acknowledgements}
\label{s:acknowledgements}

We thank Gurtina Besla, Haley Bowden, Jane Bright, Katie Chamberlain, Jaclyn Champagne, Arjun Dey, Richard Green, Jenny Greene, Kate Grier, Raphael Hviding, Takuma Izumi, Tod Lauer, Junyao Li, Jianwei Lyu, Joan Najita, George Rieke, Marcia Rieke, John Silverman, Fengwu Sun, Wei-Leong Tee, Feige Wang, Ben Weiner, Christina Williams, Charity Woodrum for very valuable discussions. This research has made extensive use of the arXiv and NASA's Astrophysics Data System. This research used the Ocelote supercomputer of the University of Arizona. The allocation of computer time from the UA Research
Computing High Performance Computing (HPC) at the University
of Arizona is gratefully acknowledged. The Bolshoi-Planck simulation was performed by Anatoly Klypin within the Bolshoi project of the University of California High-Performance AstroComputing Center (UC-HiPACC; PI Joel Primack).

%%%%%%%%%%%%%%%%%%%%%%%%%%%%%%%%%%%%%%%%%%%%%%%%%%

% Don't change these lines
\bsp	% typesetting comment
\label{lastpage}
\vspace{-0.8cm}
{\footnotesize
\bibliography{trinity}
}

\appendix

\end{document}